\documentstyle[12pt,titlepage]{article}
\begin{document}

\title{{\Large\bf Parallel Linear General
Relativity and CMB Anisotropies} \\ ~ \\
{\large\bf A Technical Paper submitted to
Supercomputing '95 in HTML format} \\ ~ \\}

\author{{\Large\bf Paul W. Bode} \\
        {bode@alcor.mit.edu} \\ ~ \\
{\Large\bf Edmund Bertschinger} \\
        {edbert@arcturus.mit.edu} \\ ~ \\
{\large\bf Massachusetts Institute of Technology} \\
{\large\bf Cambridge, MA ~~~02139} }

\date{\large\bf April 1995}

\maketitle

\begin{abstract}

This is a latex version containing most of an HTML-format
technical paper submitted for Supercomputing '95.
The preferred method of viewing is to point your WWW client
to
\begin{verbatim}
http://arcturus.mit.edu/SC95/
\end{verbatim}

We have developed a code which
links the primeval fluctuations in the early universe
with those observable at the present time
by integrating  the coupled, linearized, Einstein, Boltzmann, and
fluid equations governing the evolution of metric perturbations and
the density fluctuations;
this is the most accurate treatment to date of both the
physics and the numerical integration.
The results are useful both for calculations of the
cosmic microwave background
(CMB) anisotropy and the linear power spectrum of matter fluctuations.

The serial code (LINGER)
is highly efficient on vector machines.
Furthermore, this application is perfectly suited for coarse-grained
parallelism.
A portable, parallel implementation (PLINGER)
using common message-passing libraries
(PVM, MPI, MPL, and PVMe)
has been completed; it achieves Gflop rates on current parallel
supercomputers such as the T3D and SP2.
LINGER and PLINGER will soon to be released for general use.

\em{Supercomputing time was provided by the Cornell Theory Center and
the Pittsburgh Supercomputing Center.
This work was supported by NSF grant ASC-9318185 and NASA grant
NAG5-2816, as part of the Grand Challenge Cosmology Consortium, GC3.}
\end{abstract}

\section{Introduction}
In standard "big bang" models of cosmology,
the structures seen in the universe today formed due to self-gravity;
initially small perturbations in matter and radiation were
gravitationally amplified into galaxies and clusters of galaxies.
By following how these primordial fluctuations grow in time one can
give testable predictions for the large-scale structure of the
matter distribution.

Due to the same initial perturbations,
there are slight deviations from a pure blackbody spectrum
in the Cosmic Microwave Background (CMB) blackbody radiation,
a fossil relic of the big bang.
The COBE satellite, which confirmed the isotropy and blackbody
spectrum (with temperature 2.726 K) of the CMB, also measured
these anisotropies.  Subsequent balloon-borne and ground-based
experiments have measured the amplitude of CMB anisotropies on
different angular scales
(see Steinhardt 1995 and Lubin 1994
for recent reviews).  For a given cosmological model,
one can compute the predicted amplitudes of the CMB anisotropies
as another test of the theory.

There are a variety of cosmological parameters which are not
well constrained: the Hubble constant, neutrino masses, a possible
cosmological constant, the initial perturbation spectrum, etc.
However, once the initial conditions are set, the input physics
is well understood, and since the initial perturbations are small
the subsequent evolution is linear system.
The goal of cosmological spectral codes is to compute, for a given
matter composition and
initial spectrum of perturbations, the spectrum for mass
perturbations and CMB anisotropies expected at the present time.
These predictions can serve as a discriminant of the various models.

In the past varying degrees of approximation have been made in order
to carry out the evolution (for example Peebles \& Yu 1970, Bond \& Efstathiou
1987, Holtzman 1989, Sugiyama and Gouda 1992, among others).
The code
we discuss here has a highly
accurate treatment of both the physics and the numerical integration;
we believe it is the most accurate to date.
The tradeoff for this accuracy is increased computational cost,
making the use of supercomputers necessary.

The serial and parallel versions of the code (called LINGER
and PLINGER, respectively) will soon be made
available through the
GC3 Software Archive.

\section{Equations Governing the Evolution}
A number of equations govern the growth of perturbations.
The Einstein equations give the effect of gravity, including
the rate of universal expansion.
Also needed is the Boltzmann equation, which governs
the phase space evolution of photons and neutrinos.
Baryons, electrons, and cold dark matter
follow the equations for a pressureless perfect fluid
(though the electrons and baryons are coupled to photons by Thomson
scattering).  See
Bertschinger 1995
for a detailed introduction to the linearized Einstein equations, and
Ma \& Bertschinger 1994
for a derivation of
the equations used in the LINGER code.

Other physics which is modeled includes accurate treatments of
hydrogen and helium recombination, decoupling of photons and baryons,
and Thomson scattering (including two photon polarizations and the full
angular dependences of the scattering cross section and distribution
functions).

The equations are most easily solved in $k$-space.
In Fourier space, all the  $k$ modes in the
linearized Einstein, Boltzmann, and fluid equations
evolve independently.
In addition to the Fourier transform,  there is also an
angular expansion of the phase space distributions in terms
of Legendre polynomials;
this turns the Boltzmann equations into moment hierarchies
determined at each time step.
At a given time
it is also necessary to integrate over the 3-momentum, $q$,
of the massive neutrinos.
We carry out a full integration down to the final time
without use of any free-streaming approximation.
The time integration, ending at the present, is carried out using
the standard Runge-Kutta integrator DVERK, obtained from
netlib@research.att.com.

\section{The Serial Code}
The serial code, called LINGER, is highly efficient on vector machines;
on a single  Cray C90 node it runs at 570 Mflop,
a significant fraction of the theoretical 1 Gflop peak performance.
While the message passing version discussed below will run on the C90,
it is more efficient to use Cray's Autotasking directives
to parallelize the serial code.
This has been done, so typical speeds in excess of 8 Gflop
should be possible on 16 nodes of the C90, although the
timing runs have not yet been carried out.

For runs determining the CMB anisotropy,
we desire a high degree of accuracy, with errors
$<0.1\%$ for angular degree $l<3000$.
This requires the inclusion of up to 10,000 moments $l$,  and
the integration of up to 5000 points in $k$.
Thus this problem is computationally intensive; despite getting
570 Mflop on the Cray C90, a full run still requires roughly
75 C90 CPU-hours.

\section{The Parallel Code}
One important feature of the treatment in the previous section is that
each mode characterized by a given $k$ evolves independently.
This problem is thus perfect for coarse-grained parallelization,
since each node can work on solving the equations for a
particular value of $k$ without the need to communicate
with other nodes.

Another important feature is that
for any value of $k$, the computation necessary to evolve a
given mode to the present is much larger than that required for
the initial and final message passing.  For example, with
the smallest values of
$k$ required, the CPU time is a least two minutes on an
IBM Power2 chip, while the results are gathered as a single
message of roughly 150 bytes.
(The largest $k$-values, corresponding to
smaller scales requiring  a larger number of moments $l$,
can take up to half an hour of CPU time; the message length
increases roughly in proportion to the CPU time, to a maximum
of 80 kbyte).
Thus the overhead from message passing is insignificant.

The main loop of the serial code is in $k$; the obvious
method of parallelization is to use a master/worker approach.
The message passing required is quite straightforward.  At the beginning
of a run, the master process needs to broadcast a few
quantities to all the workers,
such as the time at which to end the evolution and
the maximum number of angular moments $l$ to compute;
it then waits for a message from any worker process.
When a worker receives this information it then requests a value
of $k$ from the master,  which replies with the appropriate
value.
When the worker completes the computation, it sends an array
containing the values of interest
back to the master, which prints out these values and sends
the  next $k$ value to the worker (or, if no further work
is to be done, a message to stop).
Thus only a few basic message passing routines are required:
broadcasting to all other nodes, sending, receiving, and checking
for an incoming message (either from a particular process or from
any process), as well as the ability to tag messages.

See Appendix A to see the algorithm in
more detail.

In the parallel code, calls
to wrapper routines are made; these routines in turn  invoke the actual
message passing libraries.  The wrapper routines are provided
in a separate file, tailored to the particular library of choice.
To date, we have used PVM
(see Geist et al. 1994), MPI (see Gropp et al. 1994),
MPL, and PVMe (available from IBM).
Given the computationally intensive nature of this code,
the choice of which library to use has no effect on the efficiency
of the code and is simply a matter of which is most convenient to
the user.

The parallel code, called PLINGER,
has been run on the DEC Alpha Cluster and the
C90/T3D at the
Pittsburgh Supercomputing Center,
and the IBM SP2 at the
Cornell Theory Center.
On the SP2, MPL requires that messages be received in the order in
which they arrive, but this does not create difficulties.  On some
machines, PVM allows the master process to cohabit a particular
node along with a worker process; this is desirable because the
master process requires little CPU time compared to the workers.
Thus PVM has a slight edge on these machines.

\section{Timing of the Parallel Code}

\subsection{Flop Rates}

\paragraph{SP2}
On a single IBM Power 2 chip, about 15 times as much CPU
is needed as for a single Cray C90 node, so the serial code
runs without special optimization at 40 Mflop, or a seventh of the 266
Mflop peak performance of the Power 2.
Thus on the IBM SP2 PLINGER achieves sustained speeds
of 2.4 Gflop using 64 nodes, and 9.6 Gflop on 256 nodes.
We have recently found that the use of the MASS
library, inlining of subroutines, and higher-order loop transformations
can significantly improve the PLINGER performance on Power 2 processors,
up to 58 Mflop on a single node;  further improvements may be possible.
Thus 15 Gflop or more should be achievable on the SP2.

\paragraph{T3D}
Rather than just using the T3D,
the message passing code is most suitable for PSC's C90/T3D
heterogeneous computing environment.  The master process
resides on the C90, handling the input/output  and controlling
the T3D processes; this master process uses a negligible amount
of CPU time.  PLINGER runs at 15 Mflop on a single
T3D node, or a tenth of the theoretical peak rate
(the flop rate was found by comparison with the C90).
For 256 nodes on the T3D the total rate  is 3.7 Gflop.

\subsection{Scaling with number of Processors}

Figure 1 shows wallclock and CPU time as a function of
the number of processors for a test run on the SP2.
The filled circles show the total CPU time (as measured by
calls to etime) divided by 100.  The open squares show the
wallclock time.
The parallel efficiency,
(total CPU time)/(wallclock time x number of nodes)),
is 95% for 32 nodes; note that these tests were not done in
dedicated mode.  The 'X' show the wallclock time for
a 256-node T3D run.
The line shows the curve expected if
the wallclock time scaled exactly as the inverse of the number
of processors.

There is practically no overhead to adding more processors, so
the CPU time does not change as the number of processors is
increased.
Once the final value of $k$ has been given to a worker
process, the other nodes will no longer have any work to
do (once they have completed the $k$-value they are currently
working on).
Thus there will be a period of time at the end of each run when not
all the processors are working, so
doubling the processors does not quite halve the wallclock time.
Since larger wavenumbers require greater computation, one simple method
by which we minimized this idle time was to compute the largest
$k$ first.
For productions runs, which are much longer than these test runs,
this idle time will be less significant.

\section{Sample Results}

\subsection{Power spectrum of the anisotropies}
The anisotropies in the CMB can be characterized in terms
of multipole moments.  The two-point temperature autocorrelation
function, $C$ compares the temperatures at points in the sky separated by
some angle.  Roughly speaking, the y-axis of this plot shows the
power in the spectrum on the angular scale of the multipole $l$.

The points in Figure 2 are experimental measurements of the CMB anisotropy.
The two leftmost points are the COBE first- and second-year data,
probing an angular scale of ten degrees. The other points are from
balloon flights or ground-based experiments; these data are available
as part of the
COSAPP
software package made available by Rahul Dave
and Paul Steinhardt at the University of Pennsylvania.
The curve shows the output of a PLINGER run using standard Cold Dark
Matter initial conditions and normalized to the COBE data.
The PLINGER run took 20 hours on 64 nodes of the SP2.
Increased accuracy in the measurement and theoretical prediction
of the power spectrum will help to discriminate between cosmological
models.

Figure 3 shows a simulated sky map, analogous to the COBE sky map,
made using the output of PLINGER.  There is much greater detail here
because  this map has not been smoothed like the COBE map;
the angular resolution is one-half degree, compared to
ten degrees for COBE.  The maximum temperature differences
are +/- 200 micro-K (with the average temperature equal
to 2.726 K).

An mpeg movie in the HTML version of this paper
shows the evolution of the potential psi of the conformal
Newtonian gauge; psi plays the role of
the gravitational potential in the Newtonian limit.
The square is a comoving 100 Mpc across (1 pc = 3.3 light years).
Standard Cold Dark Matter initial conditions are used.
The movie ends shortly after recombination, at conformal time 250 Mpc
(expansion factor 1/a = 1028).  The potential oscillates at early times
due to the acoustic oscillations of the photon-baryon fluid.  These same
oscillations produce the small angular scale features in the CMB
anisotropy map shown above.

\section{Appendix A: The Message-Passing Algorithm}

\subsection{Overview}
Here we briefly outline the message passing algorithm used.

\begin{verbatim}
Main Routine
       Initialize message passing routines.
       If master, call master subroutine; else call worker subroutine
       Exit message passing routines

The Master Subroutine
       Do initialization.
       Broadcast initial data to workers.
       While there are wavenumbers not yet complete:
          receive message from worker
          print out the data
          send a wavenumber, or a message to stop

The Worker Subroutine
       Receive initial data from master.
       Ask for wavenumber from master.
       Receive from master:  next wavenumber or message to stop
       While a message to stop has not been received:
          integrate the equations
          send the results to the master
          receive from master: next wavenumber or message to stop

Message Passing Wrapper Routines
   Certain basic message passing elements are required by PLINGER.
   We have implemented  the following routines in PVM, MPL, MPI, and PVMe.

         initpass      -initialize message passing
         endpass       -exit from message passing
         mybcastreal   -send a message to all other processes
         mysendreal    -send a message to a given process
         mycheckany    -check for message of any type from any process
         mycheckone    -check for message of a given type from a given process
         mychecktid    -check for message of any type from a given process
         myrecvreal    -receive a message
\end{verbatim}

\subsection{In More Detail}

\paragraph{Tags} Each message carries a tag which reveals its function.
\begin{verbatim}
	 tag            type
	 ---    ----------------------------------
         1     -first message from master to workers
         2     -from worker; asking for a wavenumber
         3     -from master; giving worker a wavenumber to work on
         4     -from worker; giving first set of data and lmax
         5     -from worker; giving data (length = 2*lmax+8)
         6     -from master; telling worker to stop
\end{verbatim}

\paragraph{Message Passing Wrapper Routines}
Here we show what the wrapper routines look like when using
the MPI library.

\begin{verbatim}
initpass       -initialize message passing
C   Returns process ID in mytid and the ID of the master in mastid.
       SUBROUTINE initpass(mytid, mastid)
        call MPI_INIT( ierr )
        call MPI_COMM_RANK( MPI_COMM_WORLD, mytid, ierr )
        call MPI_COMM_SIZE( MPI_COMM_WORLD, nproc, ierr )
        mastid = 0


endpass        -exit from message passing
C   Exits MPI
          SUBROUTINE endpass()
          call MPI_FINALIZE(ierr)

mybcastreal    -send a message to all other processes
C   The master process sends a message with tag=msgtype to all other processes.
C   Sends length double precision numbers starting at position buffer.
          SUBROUTINE mybcastreal( buffer, length, msgtype)
	  DO i=1, nproc-1
  	     call MPI_SEND(buffer, length, MPI_DOUBLE_PRECISION,
                                 i, msgtype, MPI_COMM_WORLD, ierr)
	  END DO

mysendreal     -send a message to a given process
C   Sends a message with tag=msgtype to the process with ID=target.
C   Sends length double precision numbers starting at position buffer.
            SUBROUTINE mysendreal( buffer, length, msgtype, target)
  	     call MPI_SEND(buffer, length, MPI_DOUBLE_PRECISION,
                            target, msgtype, MPI_COMM_WORLD, ierr)

mycheckany     -check for message of any type from any process
C   Waits for a message of any type from any process.
            SUBROUTINE mycheckany( msgtype, target)
            call MPI_PROBE( MPI_ANY_SOURCE, MPI_ANY_TAG,
                                      MPI_COMM_WORLD, status, ierr)
            msgtype = status(MPI_TAG)
	    target  = status(MPI_SOURCE)

mycheckone     -check for message of a given type from
a given process
C   Waits for a message of type msgtype from process target.
            SUBROUTINE mycheckone( msgtype, target)
            call MPI_PROBE( target, msgtype, MPI_COMM_WORLD, status, ierr)

mychecktid     -check for message of any type from a
given process
C   Waits for a message of any type from process target.
C   Returns the message tag in msgtype.
            SUBROUTINE mychecktid( msgtype, target)
            call MPI_PROBE( target, MPI_ANY_TAG, MPI_COMM_WORLD, status, ierr)
            msgtype = status(MPI_TAG)

myrecvreal     -receive a message
SUBROUTINE myrecvreal( buffer, length, msgtype, target)
   Receives a message of type msgtype from process target.
   length double precision numbers are copied, starting at address of buffer.
	    call MPI_RECV(buffer, length, MPI_DOUBLE_PRECISION, target,
     &           msgtype, MPI_COMM_WORLD, status, ierr)
\end{verbatim}

\paragraph{The Main Routine}
\begin{verbatim}
	PROGRAM plinger
C mytid = the process ID
C mastid = the ID of the master process
	INTEGER mytid,mastid
C initialize message passing routines
        CALL initpass(mytid, mastid)
	IF( mytid.EQ.mastid ) THEN
	   CALL parentsub(mytid, mastid)
	ELSE
	   CALL kidsub(mytid, mastid)
	ENDIF
C exit message passing routines
        CALL endpass
	STOP
	END
\end{verbatim}

\paragraph{The Master Subroutine}
\begin{verbatim}
	SUBROUTINE parentsub(mytid, mastid)

{The master initializes various quantities: the number of k values,
the maximum value of k, etc.  The values needed by the workers
are placed into the array y.}

C broadcast data to all node programs
	msgtype  = 1
	imsglen = 5
	CALL mybcastreal( y, imsglen, msgtype)

C ik is the next wavenumber to send
        ik = 1
C keep track of how many ik have been received from workers
	ikdone = 0

C Start checking for messages from the workers
 100    CONTINUE
        CALL mycheckany( msgtype, itid)

        IF( msgtype.EQ.2 ) THEN
C msgtype=2: the worker is ready for its first ik.
C            dispose of the message since it contains no data.
           CALL myrecvreal( deltat, 1, msgtype, itid)
        ENDIF

        IF( msgtype.EQ.4 ) THEN
C msgtype=4: receive first part of data from worker
C            the length of the next message depends on lmax
            CALL myrecvreal( y, 21, msgtype, itid)
	    ikold = INT(y(1))
	    lmax = INT(y(21))
C this data is written to an ascii file
	    WRITE(unit_1,*) (y(i),i=1,20)

C msgtype=5: receive second part of data from worker
            msgtype = 5
            CALL mycheckone( msgtype, itid)
            CALL myrecvreal( y, 8+lmax+lmax, msgtype, itid)
C this data is written to a binary file
	    WRITE(unit_2) ikold,(y(i),i=7,lmax)
	    WRITE(unit_2) (y(8+lmax+i),i=0,lmax)
C done with this wavenumber
	    ikdone = ikdone+1
	ENDIF

        IF( (msgtype.EQ.2).OR.(msgtype.EQ.5) ) THEN
	    IF( ik.LE.last_nk) THEN
C reply with ik to the worker that sent the last message
               y(1) =  DBLE(ik)
	       msgtype  = 3
	       imsglen  = 1
	       CALL mysendreal( y, imsglen, msgtype, itid)
C find the next value of ik to be sent
               CALL ik_next(ik)
            ELSE
C if no more wavenumbers to send, tell the worker to stop (msgtype=6)
	       msgtype  = 6
	       imsglen  = 1
	       CALL mysendreal( y, imsglen, msgtype, itid)
	    ENDIF
        ENDIF

C if there are still wavenumbers to receive, go back
	IF( ikdone.LE.last_nk) GO TO 100

	RETURN
	END
\end{verbatim}

\paragraph{The Worker Subroutine}
\begin{verbatim}
	SUBROUTINE kidsub(mytid, mastid)

C receive initial data from master
        msgtype  = 1
        CALL mycheckone( msgtype, mastid)
	CALL myrecvreal( passbuffer, 5, msgtype, mastid)

C ask for wavenumber from master
        msgtype  = 2
	imsglen = 1
	CALL mysendreal( passbuffer, imsglen, msgtype, mastid)

C receive from master:  next ik or message to stop
        CALL mychecktid( msgtype, mastid)
	CALL myrecvreal( passbuffer, 1, msgtype, mastid)
	ik = INT( passbuffer(1) )

C if the message is not a wavenumber, then exit
        WHILE ( msgtype.NE.3 )

C  Begin timestep loop.
            time = t_start
            WHILE( time.LT.end_time)

{Here the worker integrates the coupled equations.} </I> <BR>

	    END WHILE

C send first part of results to master
            msgtype  = 4
	    imsglen = 21
	    CALL mysendreal( passbuffer, imsglen, msgtype, mastid)
C send second part of results to master
            msgtype  = 5
            imsglen = 8+lmax+lmax
            CALL mysendreal( y, imsglen, msgtype, mastid)

C receive from master:  next ik or message to stop
            CALL mychecktid( msgtype, mastid)
            CALL myrecvreal( passbuffer, 1, msgtype, mastid)
            ik = INT( passbuffer(1) )

	END WHILE

       RETURN
       END
\end{verbatim}

\section{References}

\subsection{Articles}

\begin{list}{}{\leftmargin .75in \itemindent -0.5in \itemsep 0in}

\item  Lubin, P.M. 1994, in {\it Examining the Big Bang and the
Diffuse Background Radiations: Proceedings of IAU Symp. 168}
(astro-ph/9412021)

\item   Ma, C., and E. Bertschinger 1994, preprint
(astro-ph/9401007)

\item   Bennett, C.L. et al. 1994, Ap. J. 436 423

\item   Bertschinger, E. 1995, in {\it 1993 Les Houches Summer School
Lectures on Cosmology} (Elsevier Science Publishers B.V.)
(astro-ph/9503125)

\item   Bond, J.R., and G. Efstathiou 1987 MNRAS 226 655

\item   Geist, A., et al. 1994, {\it PVM: Parallel Virtual Machine}
(MIT Press, Cambridge MA)
(HTML version available at
\begin{verbatim}
http://www.netlib.org/pvm3/book/pvm-book.html)
\end{verbatim}

\item   Gropp, W., E. Lusk, and A Skjellum 1994, {\it  Using MPI}
(MIT Press, Cambridge MA)

\item   Holtzman, J. 1989, Ap. J. Supp. 71 1

\item   Peebles, P.J.E., and J.T. Yu 1970 Ap. J. 147 73

\item   Smoot, G.F. et al. 1992, Ap. J. Lett. 396 L1

\item   Steinhardt, P.J. 1995, in {\it Snowmass Workshop on Particle
Astrophysics and Cosmology}, ed. E. Kolb and R.Peccei
(astro-ph/9502024)

\item   Sugiyama, N., and N. Gouda 1992, Prog. Theor. Phys. 88 803

\end{list}

\subsection{Web links}

{\bf Message passing libraries:}
\begin{verbatim}
http://www.epm.ornl.gov/pvm/pvm_home.html
        PVM: Parallel Virtual Machine

http://www.mcs.anl.gov/Projects/mpi/index.html
        MPI - Message Passing Interface

http://www.mcs.anl.gov/Projects/mpi/index.html
        MPI - Message Passing Interface

http://ibm.tc.cornell.edu/ibm/pps/doc/pvme.html
        IBM POWERparallel Systems: PVMe
\end{verbatim}

{\it See the Cornell Theory Center Hardware Page for information on MPL} \\

\noindent{\bf Supercomputing Centers (for information on the machines used):}
\begin{verbatim}
http://www.tc.cornell.edu/UserDoc/Hardware/
        Cornell Theory Center Hardware Page

http://pscinfo.psc.edu/
        Pittsburgh Supercomputing Center home page
        (CMB Anisotropy Experiment Data, compiled by Rahul Dave)
\end{verbatim}

\noindent{\bf Astrophysics Servers:}
\begin{verbatim}
http://dept.physics.upenn.edu/~www/astro-cosmo
        CMB Window and Bandpower Software Package (COSAPP)

http://xxx.lanl.gov/astro-ph/
        Astro-ph Astrophysics Preprints
\end{verbatim}

\noindent{\bf The Grand Challenge Cosmology Consortium, GC3:}
{\it Look for LINGER (as part of the COSMICS cosmological initial
conditions package) and PLINGER to be made available from these sites.}
\begin{verbatim}
http://zeus.ncsa.uiuc.edu:8080/GC3_Home_Page.html
        The GC3 Home Page

http://zeus.ncsa.uiuc.edu:8080/GC3_software_archive.html
        The GC3 Software Archive

http://arcturus.mit.edu/GC3/
        The GC3 - MIT Branch Home Page

http://arcturus.mit.edu/~bode/
        Paul Bode's Home Page
\end{verbatim}

\section{Figures}

\paragraph{1. } Wallclock and CPU time as a function of
the number of processors for a test run on the SP2.

\paragraph{2. } The points are experimental measurements of the CMB anisotropy,
from the COSAPP package.
The curve shows the output of a PLINGER run using standard Cold Dark
Matter initial conditions and normalized to the COBE $Q_{rms-PS}$.

\paragraph{3. } A simulated sky map, analogous to the COBE sky map,
made using the output of PLINGER.

\end{document}